\documentclass[useAMS,usenatbib]{mn2e}
\usepackage{graphicx,natbib,amsmath,color}
\usepackage[T1]{fontenc}
\usepackage{aecompl}

\newcommand{\comment}[1]{}
\newcommand{\changed}[1]{{#1}}

\title[Optical turbulence profiling using Stereo--SCIDAR]{Stereo--SCIDAR: Optical turbulence profiling with high sensitivity using a modified SCIDAR instrument}

\author[H.~W. Shepherd et al.]{H.~W. Shepherd$^{1}$, J. Osborn$^{1}$\thanks{E-mail:
james.osborn@durham.ac.uk (JO)}, R.~W. Wilson$^{1}$, T. Butterley$^{1}$, R. Avila$^{2}$\\ 
\newauthor{V.~S. Dhillon$^{3}$ and T.~J. Morris$^{1}$} \\
$^{1}$ 
Department of Physics, Centre for Advanced Instrumentation, University of Durham, South Road, Durham DH1 3LE, UK \\
$^{2}$
Centro de Fisica Aplicada y Technolg\`{i}a Avanzada, Universidad Nacional Aut\`{o}noma de M\`{e}xico, A.P. 1-1010, Santiago de Quer\`{e}taro, \\ Quer\`{e}taro, 76000, M\`{e}xico\\
$^{3}$
Department  of Physics and Astronomy, University of Sheffield, Sheffield, S3 7RH}
\begin{document}


\maketitle
\label{firstpage}

\begin{abstract}
The next generation of adaptive optics (AO) systems will require tomographic reconstruction techniques to map the optical refractive index fluctuations, generated by the atmospheric turbulence, along the line of sight to the astronomical target. These systems can be enhanced with data from an external atmospheric profiler. This is important for Extremely Large Telescope scale tomography. Here we propose a new instrument which utilises the generalised SCIntillation Detection And Ranging (SCIDAR) technique to allow high sensitivity vertical profiles of the atmospheric optical turbulence and wind velocity profile above astronomical observatories. The new approach, which we refer to as `Stereo--SCIDAR', uses a stereoscopic system with the scintillation pattern from each star of a double-star target incident on a separate detector. Separating the pupil images for each star has several advantages including: increased magnitude difference tolerance for the target stars; negating the need for re-calibration due to the normalisation errors usually associated with SCIDAR; an increase of at least a factor of two in the signal-to-noise ratio of the cross--covariance function and hence the profile for equal magnitude target stars and up to a factor of 16 improvement for targets of 3 magnitudes difference; and easier real-time reconstruction of the wind-velocity profile. Theoretical response functions are calculated for the instrument, and the performance is investigated using a Monte-Carlo simulation. The technique is demonstrated using data recorded at the 2.5~m Nordic Optical Telescope and the 1.0~m Jacobus Kapteyn Telescope, both on La Palma. 
\end{abstract}


\section{Introduction and Theory}
\changed{
Several techniques have been implemented to estimate the atmospheric turbulence profile, measured as the refractive index structure constant, $C_n^2\left(h\right)$, and wind velocity, both as a function of altitude. The most widely exploited are MASS (Multi Aperture Scintillation System, \citealp{Tokovinin07}), SCIDAR (SCIntillation Detection And Ranging, \citealp{Vernin73}) and SLODAR (SLOpe Detection And Ranging, \citealp{Wilson02}). MASS is not intended as a high vertical-resolution technique. It has a limited logarithmic vertical resolution and the high altitude response is very broad \citep{Tokovinin07}. Here, we only address high altitude-resolution techniques and therefore we will only discuss SLODAR and SCIDAR. Both SLODAR and SCIDAR are triangulation techniques in which the atmospheric turbulence profile is recovered from either the correlation of wavefront slopes in the case of SLODAR, or scintillation intensity patterns in the case of SCIDAR, for two target stars with a known angular separation. A simplified schematic is shown in figure~\ref{fig:triangulation}.
\begin{figure}
	\centering
	\includegraphics[width=0.35\textwidth]{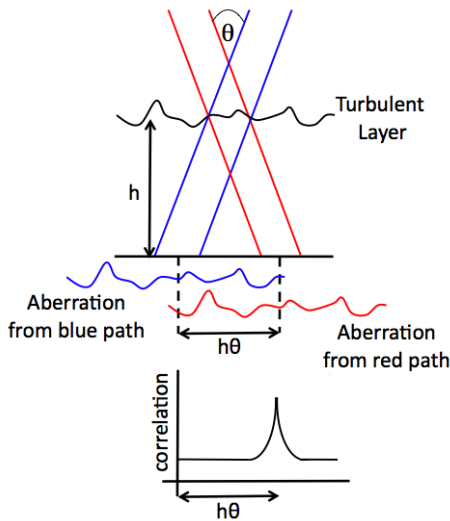}
	\caption{If a turbulent layer at height, $h$, is illuminated by two stars of angular separation, $\theta$, then two copies of the aberration will be made on the ground separated by a distance $h\theta$. By cross correlating either the centroid positions from a Shack--Hartmann wavefront sensor (SLODAR) or the intensity patterns (SCIDAR) we can triangulate the height of the turbulent layer and the amplitude of the correlation peak corresponds to the strength of the layer.}
	\protect\label{fig:triangulation}
\end{figure}

For the next generation of Extremely Large Telescopes (ELT) the knowledge of the turbulence profile will become essential to the efficient running of the observatory and adaptive optics (AO) systems. Tomographic AO systems combine information from several off-axis wavefront sensors to estimate the optical phase aberrations in the volume of turbulence in a given target direction or field of regard. These multi-wavefront sensor systems can build an atmospheric profile using the internal wavefront sensor data using the SLODAR method \citep{Cortes12}. External profilers can provide information which is always of the highest altitude resolution irrespective of the ELT AO wavefront sensor configuration and is independent of the AO system.

Externally generated profiles can be used to augment information derived from internal WFS data and assist observatory operations. These tasks could include validation of the performance of the AO instrumentation; comparison of performance with calculated error budgets; diagnosis of system performance issues at an early stage; collection of site statistics; validation and feedback into meteorological forecasting models \citep{Masciadri12}; pre-optimisation of AO control matrices to minimise telescope down-time between observations; construction of AO control matrices for fields without any bright targets and potentially the scheduling of AO observations. However, the precise role of external profilers in the ELT era is still an area of active research and is outside the intended scope of this paper.

Turbulence profilers such as SLODAR or SCIDAR can also supply wind velocity measurements which will allow the tomographic reconstructors to use temporal information in the reconstruction process. Several smart reconstructors (for example, Linear Quadratic Control, \citealp{Folcher11}) require real time wind velocity profiles to optimise the AO control algorithms. The combined atmospheric turbulence strength and velocity profile can be used to calculate atmospheric parameters important for the real time optimisation of AO systems, such as the isoplanatic angle and coherence time.
}

In this paper we discuss the SCIDAR technique. SCIDAR has the capability to determine the atmospheric profile to a higher resolution than SLODAR because the auto--covariance function of the scintillation pattern from a single turbulent layer is narrower than that of the wavefront slopes, i.e. the spatial scale of the scintillation is smaller than the minimum wavefront sensor sampling of the phase, allowing for higher altitude resolution profiling. SCIDAR was originally proposed by \cite{Vernin73}, in which the turbulence profile is determined by processing short exposure images of the scintillation pattern observed from a double star. SCIDAR is limited in that it is insensitive to turbulence at the\changed{ ground, due to lack of propagation distance required to develop scintillation}. \cite{Fuchs94} introduced generalised--SCIDAR with the suggestion that the analysis plane did not necessarily need to be at the telescope pupil. A conjugate position larger than 1~km below the pupil was suggested by \cite{Fuchs98} and even larger distances were used in the first implementation of generalised--SCIDAR by \cite{Avila97}. This extends the propagation distance of the light path and therefore allows the phase aberrations induced by the surface turbulent layer to develop into intensity fluctuations, which can be measured.

The theoretical resolution for SCIDAR \changed{is defined by} the Fresnel zone size for a given altitude of the turbulent layer and is given by \citep{Prieur01},
\begin{equation}
	\delta h\left(z\right) = 0.78\frac{\sqrt{\lambda z}}{\theta},  
	\label{eq_dh}
\end{equation}
where $z$ is the propagation distance to the layer and is given by $z=| h - h_{\mathrm{conj}} |$, where $h_{\mathrm{conj}}$ is the conjugate altitude of the detector, or analysis plane, and $h$ is the altitude of the turbulent, $\lambda$ is the wavelength and $\theta$ is the angular separation of the target stars. \changed{Throughout this paper the analysis assumes a zenith angle, $\theta_z =0$. To generalise to other zenith angles, one would replace $z$ by $z\sec{\theta_z}$.} The altitude resolution for SCIDAR is a function of the propagation distance and target star separation.\comment{ and is shown in figure \ref{fig:scidarRes}.} For larger propagation distances the spatial scale of the intensity speckle patterns is larger, reducing the altitude resolution. Stars with a wider separation will increase the altitude resolution but also reduce the maximum profiling altitude\changed{, as we can only recover the profile up to an altitude where the projected pupils from the two stars overlap. The maximum altitude that we can observe a layer is therefore,}
\begin{equation}
 h_{\mathrm{max}}=\frac{D}{ \theta}, 
 \label{eq:hmax}
\end{equation}
where $D$ is the diameter of the telescope pupil.\changed{ The maximum altitude will actually be lower than this in the case where the analysis plane is positioned away from the telescope pupil. In this case diffraction through the pupil will distort the intensity distribution at the edge of the pupil (outer, secondary and spiders) and will need to be blocked. This will effectively reduce the diameter of the telescope. The width of the primary diffraction ring is independent of telescope size and is given by the Fresnel radius, $r_F=\sqrt{\lambda h_\mathrm{conj}}$. This ring is substantially larger than the others and so only this outer one is blocked in order to retain a large fraction of the telescope size \citep{Osborn11}. Equation~\ref{eq:hmax} is modified to,
\begin{equation}
 h_{\mathrm{max}}=\frac{(D-r_F)}{ \theta}. 
 \label{eq:hmax2}
\end{equation}
}

For a given telescope, $\theta$ should be selected so that $h_{\mathrm{max}}$ is approximately 20~km, the maximum expected altitude of the tropopause and hence the maximum altitude for any \comment{significant} optical turbulence.

To improve the sensitivity, efficiency and resolution of generalised--SCIDAR we have designed and built a new instrument with two cameras, one for each target star, named stereo--SCIDAR. Separating the intensity patterns from each star on to two independent cameras  \changed{increases the signal to noise ratio of the profile, allows a greater magnitude difference of the targets} and circumvents known normalisation issues with conventional single camera generalised--SCIDAR data reduction \citep{Avila09}.

Section 2 describes the Stereo--SCIDAR technique including the opto-mechanical design, the theoretical derivation of the response functions and an overview of the associated advantages. Section 3 explains the data reduction algorithm and the profile fitting. Section 4 presents a selection of on-sky results. \comment{Section 6 discusses possible observing strategies to optimise the output depending on the requirements of the application and the implications of probing rather than profiling. }We conclude in section 5.

\section{Stereo--SCIDAR}

\subsection{Opto--Mechanical Design}

\begin{figure}
	\centering
	\includegraphics[width=0.4\textwidth]{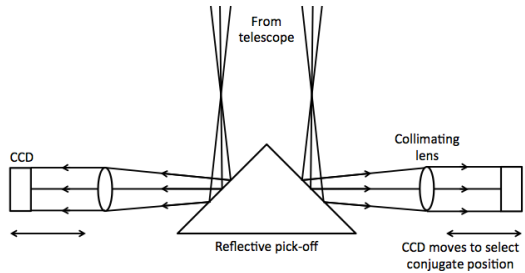}
	\caption{A schematic layout of the Stereo SCIDAR instrument. The reflecting pickoff in the centre directs the light from each star to one arm of the instrument. The light in each arm is then collimated, and the resulting intensity pattern is imaged onto the CCD detector. Each CCD detector is attached to a linear stage which controls the altitude of the alaysis plane imaged by the detector.}
	\protect\label{fig:diagram}
\end{figure}

Conventionally, SCIDAR is implemented with a single camera, which records the overlapping pupil images. Stereo--SCIDAR utilises a separate camera for each pupil image. The prototype Stereo--SCIDAR instrument has been deployed on the 2.5~m Nordic Optical Telescope (NOT) during March 2013 and the 1~m Jacobus Kapteyn Telescope (JKT) in May, July and September 2013. The instrument uses a reflective glass wedge to separate the beams from each of the two stars onto two separate CCD detectors. One of the images is inverted in software to ensure that both images have the same on--sky orientation. A sketch of the instrumental design is shown in figure \ref{fig:diagram}. In this implementation two Andor Luca S EMCCD detectors were used. \changed{The cameras have a maximum frame rate of $\sim$90Hz and we use an exposure time of 2~ms.} \comment{The cameras have an approximate quantum efficiency of 40\% between 400 and 750~nm.}The NOT has an aperture diameter of 2.5~m and an effective focal length of 28.2~m. A collimating lens of focal length 32~mm leads to a beam diameter of 2.8~mm. \comment{ The magnification of this system $G$ is 880, and the shift in conjugate altitude $\Delta h$ is related to the shift in detector position $\Delta s$ by,
\begin{equation}
	\Delta h = G^2 \Delta s.
\end{equation}
Hence a shift of the detector by 1.3~mm along the optical axis leads to a 1~km change in conjugate altitude. When the system is deployed on the JKT we use a Barlow lens to increase the focal length of the optical system, which results in an altitude scale of 1.0~mm in conjugate space corresponding to 1~km of propagation.}

A two camera system can emulate any functionality of a single camera system by simply adding together the two images. However, there are several advantages to the two camera system at the cost of a more complicated opto-mechanical and electronic design.

\comment{
\subsection{Optical Conjugation}
Figure~\ref{array:recon} shows how reconjugation can be produced in the optical system. Consider the simplified case of a single layer of turbulence at a high altitude. This produces scintillation in the entrance pupil of the telescope. If we reimage closer to the high altitude layer then the rays will have propagated so as to partially ``undo'' some of the scintillation and we would be left with a wavefront which appears to have propagated a smaller distance \citep{Fuchs98,Osborn11}.

\begin{figure}
	\centering
	 \includegraphics[width=0.5\textwidth]{recon_diag3.eps} \\
	 \caption{Ray diagrams for conjugation positions. The black lines show the rays for an object at infinity. The top diagram shows the conjugate position of the telescope pupil. Every point in this plane will be an image of a point on the telescope pupil (as shown by the dashed lines). The lower diagram shows that by moving the observation plane towards the collimating lens we can measure the intensity distribution at a height $h$ above the telescope. If a camera is in a position such that it is in the image plane of the turbulent layer it is at the conjugate altitude of that layer. In practice subsidiary optics may also be used, but this diagram shows the principle.}
	 \label{array:recon}
\end{figure}
}

\subsection{Cross--covariance functions}

\begin{figure}
	\centering
	 \includegraphics[height=0.45\textwidth,angle=90]{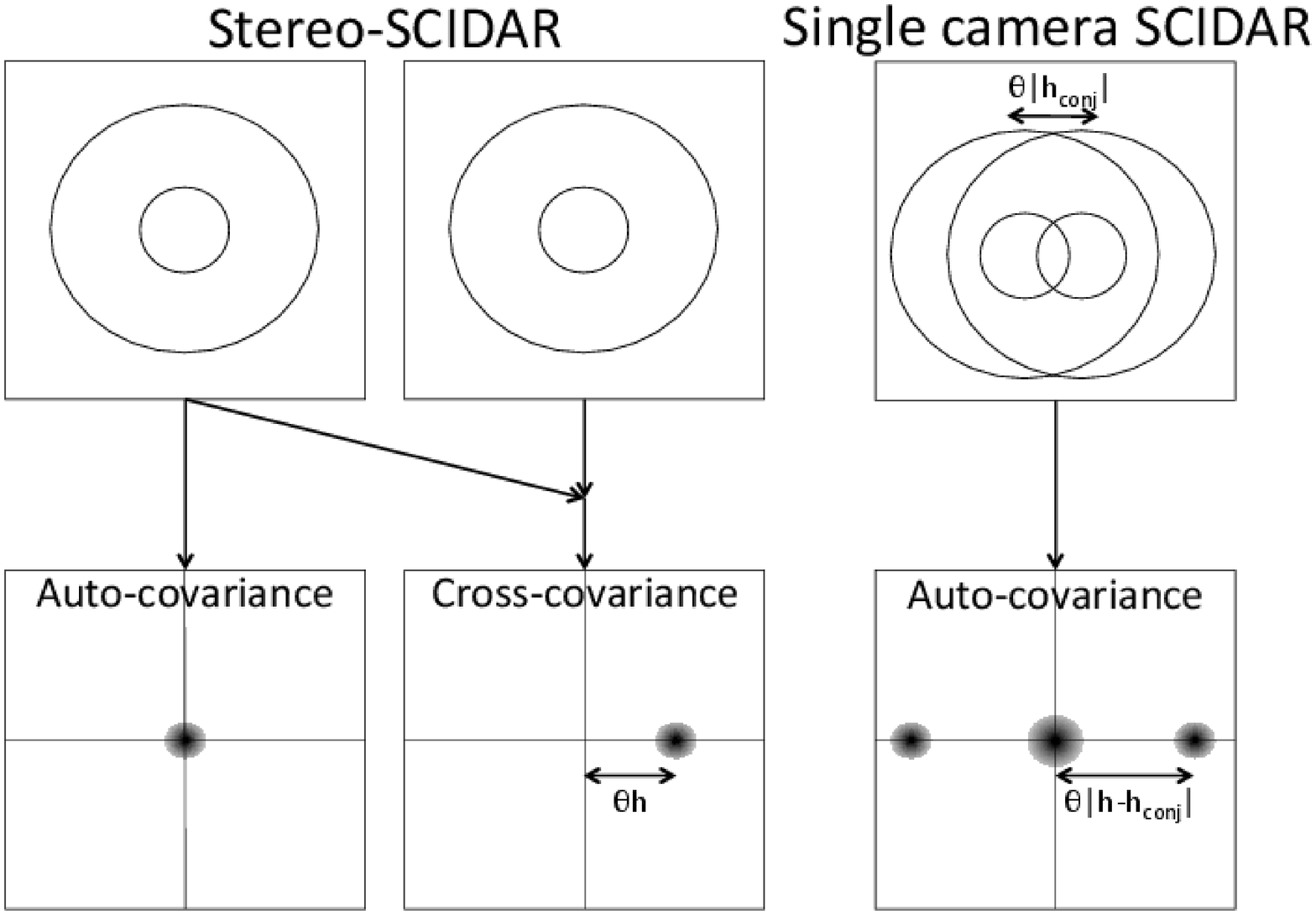} \\
	 \caption{The cross--covariance function for the two stereo--SCIDAR frames results in a single peak offset by $\theta h$ (left). The auto-covariance function of single camera SCIDAR results in three peaks for each layer. One at the centre of the auto--covariance and one at plus and minus $\theta |h-h_{\mathrm{conj}}|$. (right)}
	 \label{fig:auto_cross}
\end{figure}

Using two cameras instead of one changes the profile restoration process for SCIDAR. When using a single camera, the pupil images from each target star overlap on the CCD, with an offset depending on the conjugate altitude of the analysis plane and on the angular separation of the targets (at the telescope pupil the two images will be completely superimposed). One calculates the auto--covariance function of the image which is normalised by the auto--covariance of the mean image. Each turbulent layer will contribute three peaks to the auto--covariance function, as the intensity pattern from each star correlates with the intensity pattern with the other star and with itself (figure~\ref{fig:auto_cross}, right). The profile can be restored by fitting the 1D cut of the covariance peaks in the direction between the two stars with the theoretical response functions of the instrument. The response functions map the atmospheric optical turbulence profile onto the covariance function (i.e what the instrument actually measures).

\begin{figure}
\centering
\includegraphics[width=0.5\textwidth]{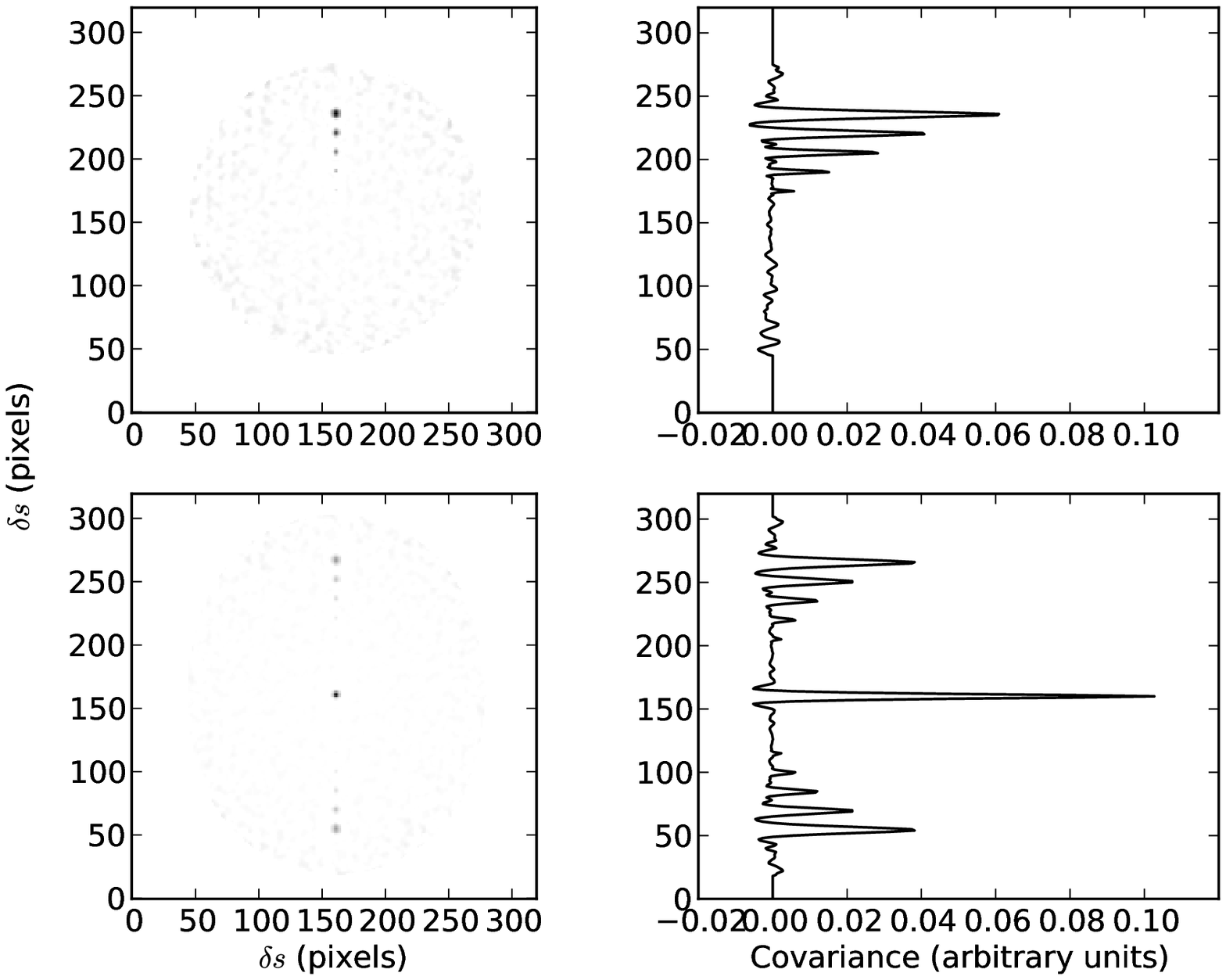}
\caption{2D covariance plots for Stereo--SCIDAR (upper) and single camera Generalised--SCIDAR (lower) for an atmospheric simulation containing six equal strength turbulent layers at 2~km spacing between 0 and 10~km, inclusive. A  vertical cut through each covariance function is shown on the right. $\delta s$ is the position in the covariance function. We see that for single camera SCIDAR we have two sets of spatially separated peaks and one set of overlapping peaks at the centre. For Stereo--SCIDAR we only have one set. Both plots have the same grey scale, the correlation peaks for Stereo-SCIDAR are larger in magnitude than that of single camera Generalised--SCIDAR.}
\label{fig:xcov2cam}
\end{figure}

For a two camera system we normalise each pupil image to have the same mean intensity and then calculate the cross--covariance of the two individual intensity patterns. This cross--covariance can be normalised by the cross--covariance of the mean pupil images. In this way the Stereo--SCIDAR cross--covariance function only has one set of correlation peaks from the centre outwards in the direction parallel to the position angle of the two binary target stars (figure~\ref{fig:auto_cross}, left). Figure~\ref{fig:xcov2cam} shows example simulated covariance functions for Stereo--SCIDAR and for single camera SCIDAR.

The spatio-temporal covariance (cross or auto) can be computed by calculating the covariance function with increasing offsets in the frame number. If we correlate one frame with itself, this gives us the $dt=0$ \changed{plane}. The correlation of one frame with the subsequent frame would be the $dt=1$ \changed{plane} and with the preceding frame would be the $dt=-1$ \changed{plane} in the spatio-temporal covariance.

\subsection{Theoretical Background}

\subsubsection{Kolmogorov Turbulence}
\label{sect:subtheory}

The propagation of starlight through a turbulent layer in the atmosphere leads to an intensity distribution at the telescope aperture. If we assume that the turbulence has a Kolmogorov power spectrum then the spatial intensity power spectrum at the ground is given by,
\begin{gather}
	\Phi_{I}\left(f\right) = 3.9\times10^{-2} k^2 f^{-11/3}  \nonumber \\
	 \int_0^\infty C_n^2\left(z\right) \sin^2\left(\pi\lambda z f^2\right) \rmn{d}h,
	\protect\label{eqn:scint_spectrum}
\end{gather}
where $k$ is the wave-number of the light from the star, and $f$ the spatial frequency of the atmospheric turbulence \citep{Roddier1981}. For Kolmogorov turbulence this is valid over the inertial range $1/L_0 \leq f \leq 1/l_0$, where $L_0$ and $l_0$ are the outer and inner scales of the turbulence respectively. 

\changed{
The above equation forms the basis of the SCIDAR response functions. It should be noted that it is only valid for weak-scintillation conditions and in the monochromatic light approximation. In SCIDAR techniques, scintillation only occasionally fails to be weak. In such cases, the power spectrum of the actual scintillation has lower maximum values and presents a noisy aspect, as shown by \cite{Tokovinin07} from numerical simulations. Essentially, in the strong fluctuation regime, the effect of each independent turbulent layer can no longer be assumed to be additive. 

The validity of the monochromatic approximation has been well justified for SCIDAR-like techniques \citep{Tokovinin03}. Here we use polychromatic light, only filtered by the wavelength dependent quantum efficiency of our camera. All SCIDAR systems use polychromatic light, in order to collect sufficient photons for the analysis. We have developed a Monte-Carlo simulation to compare the monochromatic covariance function with a polychromatic one. The simulation included a test atmosphere containing six layers separated by 2~km and all having equal turbulence strength, this was to confirm that the effect of several turbulent layers was additive (weak fluctuation regime), as is assumed by SCIDAR. The polychromatic pupil images were generated by summing together six images generated with wavelengths between 500 and 800~nm, outside of this range the quantum efficiency of the cameras ensure that the throughput is negligible. The images were weighted by the wavelength dependent quantum efficiency of the camera and the stellar spectral emission. The transmission spectrum of the atmosphere in the visible was assumed to be uniform. The wavelength weights are as shown in table 1.
\begin{table*}
\centering
\caption{Wavelength weights for polychromatic simulation}
\begin{tabular}{ccccc}
\hline
Wavelength (nm) & Andor Luca QE \% & Stellar flux / maximum & Weight\\
\hline
500 & 0.52  & 1.00 & 0.21\\
550 & 0.5    & 0.92 & 0.19\\
600 & 0.49  & 0.86 & 0.17\\
650 & 0.45  & 0.78 & 0.15\\
700 & 0.38  & 0.77 & 0.12\\
750 & 0.29  & 0.75 & 0.09\\
800 & 0.22  & 0.73 & 0.07\\
\hline
\end{tabular}
\end{table*}

We compared the simulated polychromatic covariance functions with the monochromatic case (at $\lambda=500$~nm). The simulation showed negligible difference between the two covariance functions meaning that it is indeed appropriate to use the monochromatic equations to generate the response functions of SCIDAR.\comment{ (figure~\ref{fig:polychromo}).
\begin{figure}
\centering
\includegraphics[width=0.5\textwidth]{polychromoSim}
\caption{Simulated covariance functions for monochromatic and polychromatic light. The difference between the two curves is negligible, implying that we can use the monochromatic response functions with polychromatic SCIDAR.}
\label{fig:polychromo}
\end{figure}
In addition to the chromatic decorrelation described above, chromatic refraction is also a problem at larger zenith angles. For this reason we restrict our observing to zenith angles less than 35~degrees.
}
}

\subsubsection{SCIDAR response functions}
The response function is the response of the instrument to a thin layer at a given altitude and maps the output of the instrument (i.e the covariance function) to the actual turbulence profile. The Stereo--SCIDAR response functions are identical to those of conventional SCIDAR, the difference between the techniques is in how the covariance function is generated and normalised.

The quantity measured by the Stereo-SCIDAR instrument is the normalised scintillation spatial auto--covariance function $B\left(\vec{r}\right)$, 
\begin{gather}
	B\left(\vec{r}\right) = \frac{\left<\left[I_1\left(\vec{x}\right)-\left<I_1\right>\right]\left[I_2\left(\vec{x}+\vec{r}\right)-\left<I_2\right>\right]\right>}{\left<I_1\right> \left<I_2\right>},
	\protect\label{eqn:normalised_scint_spat}
\end{gather}
where $I_n\left(\vec{x}\right)$ is the normalised intensity distribution in the analysis plane for a single star, $n$. The angled brackets denote an ensemble average.

As the power spectrum of the intensity variations for both stars are identical, the auto--covariance of the intensity patterns can be related to the power spectrum of the scintillation using the Wiener-Khintchine theorem. This states that the auto--covariance corresponds to the Fourier transform of the power spectrum. As the power spectrum is rotationally symmetric, this can be taken to be a Hankel transform \citep{Roddier1981}, 
\begin{equation}
	B\left(\vec{r}\right) = 2\pi\int_0^{\infty} f \Phi_I\left(f\right) J_0 \left( 2\pi r f \right) \rmn{d}f,
	\protect\label{eqn:wk_theorem}
\end{equation}
where $J_0$ is a Bessel function of the first kind. By combining equations \ref{eqn:scint_spectrum} and \ref{eqn:wk_theorem}, a single turbulent layer at propagation distance $z$ will give,
\begin{gather}
	B\left(\vec{r},z\right) = 3.9\times10^{-2} 2\pi k^2 C_{n}^{2}\left(z\right) dh\nonumber \\
					 \int_0^\infty f^{-8/3} \sin^2\left(\pi\lambda z f^2\right) J_0 \left(2\pi r f\right) \rmn{d}f.
	\protect\label{eqn:bh}
\end{gather}
This is the intensity fluctuation auto--covariance density per unit altitude produced by a layer located at a distance $z$. We also define a further quantity $B^{\prime}(\vec{r},z)$, corresponding to $B (\vec{r},z)$ for unit turbulence strength \changed{(i.e. $B(\vec{r},z)/C_n^2(h)dh$)}. We define this quantity to be the Stereo--SCIDAR response function for unit turbulence at altitude $z$. 

The measured star auto--covariance ($B\left(\vec{r},\theta\right)$) corresponds to
\begin{equation}
	B\left(\vec{r},z\right) = \int C_{n}^{2}\left(z\right) B^\prime\left(\vec{r},z\right) \rmn{d}z.
	\label{eq:sscidar_respfn}
\end{equation}
\comment{where $\theta$ is the angular separation between the stars. }

\begin{figure}
	\centering
	\includegraphics[width=0.45\textwidth]{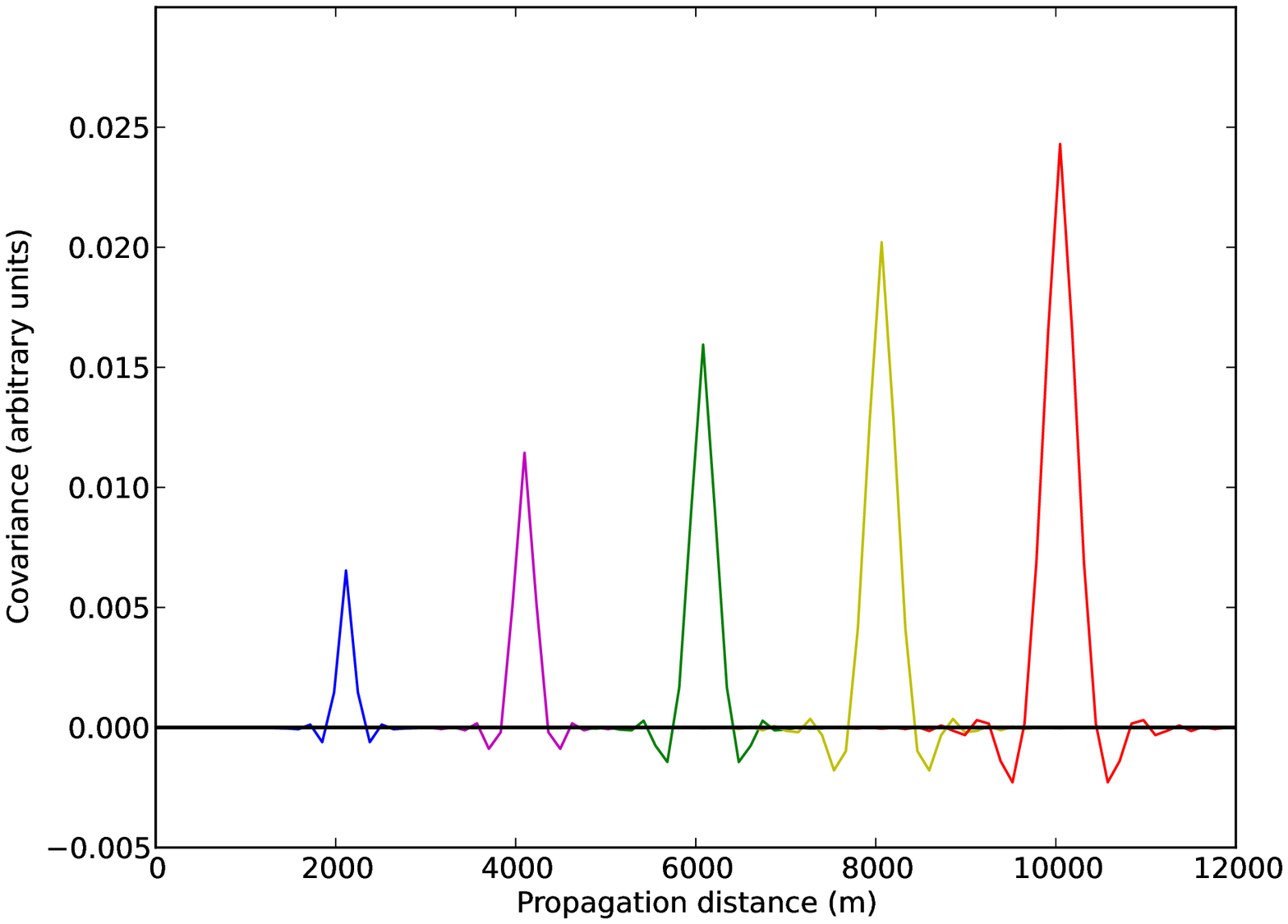}
	\caption{A sub-sample of SCIDAR response functions corresponding to five propagation distances equally spaced between 2 and 10 km. Here we plot the propagation distance and not absolute altitude and so, for example, the result is identical for a layer at 2~km whilst conjugate to 0~km and for a layer at 0~km whilst conjugate to -2~km.}
	\protect\label{fig:responseFn}
\end{figure}

Example SCIDAR response functions are shown in figure \ref{fig:responseFn} for five propagation distances. As the propagation distance increases, the cross--covariance signal increases, but also becomes wider, decreasing the altitude resolution. \comment{By dynamically reducing the propagation distance of a turbulent layer to the detector via re-conjugation, the resolving power is improved as the response function is narrower.} \comment{The stereo--SCIDAR theoretical altitude resolution is independent of the altitude of the turbulent layer, only the propagation distance from the layer altitude to the conjugate altitude is important.}

\subsubsection{Single camera SCIDAR auto-covariance functions}
\label{sect:A1A2}
In single camera SCIDAR the measured auto--covariance function can be written as \citep{Tokovinin97},
\begin{gather}
B(\vec{r},\vec{\theta}) = \int A_1 B(\vec{r})  + A_2\left(B\left(\vec{r}-\vec{\theta} z,z\right)  \right. \nonumber\\
\left. +B\left(\vec{r}+\vec{\theta} z,z\right)\right) dz,
\label{eqn:B_GS}
\end{gather}
where,
\begin{equation}
A_1= \frac{1+\gamma^2}{(1+\gamma)^2}, 
\label{eq:A1}
\end{equation}
\begin{equation}
A_2=\frac{\gamma}{(1+\gamma)^2},
\label{eq:A2}
\end{equation}
and
\begin{equation}
\gamma=10^{-0.4\Delta m},
\end{equation}
where $\Delta m$ is the relative magnitude difference of the target stars and $B(r)$ is the scintillation auto--covariance function. Each term within the integral of equation~\ref{eqn:B_GS} corresponds to a set of peaks in the auto--covariance function (figure~\ref{fig:xcov2cam}). To recover the turbulence profile we take either of the lateral sets of peaks from the triplet ($B(r-\theta |h-h_{\mathrm{conj}}|)$ or $B(r+\theta |h-h_{\mathrm{conj}}|)$) leading away from the centre and fit them to the response functions of the instrument. From equation~\ref{eqn:B_GS} we see that the amplitude of these peaks, and hence the visibility, are multiplied by $A_2$. It can be shown that the uncertainty in the optical turbulence profile $\propto A_1/A_2$ \citep{Prieur01}. Therefore, for larger magnitude differences the visibility of the correlation peaks is reduced, hence reducing the signal to noise ratio. If both stars have the same brightness then $A_1 = 0.5$ and $A_2=0.25$. However, if there is a two magnitude difference in brightness then $\gamma = 0.16$, $A_1 = 0.76$ and $A_2= 0.12$.

\subsection{Advantages of Stereo--SCIDAR}

\subsubsection{Normalisation}
In conventional single camera generalised-SCIDAR the pupil patterns from the two target stars overlap. This overlap results in a lack of contrast in the combined pupil image and hence a loss of information. The loss of contrast means that the peaks in the covariance function that are used to recover the profile are reduced in amplitude. 

If the analysis plane is conjugate to the ground, then the pupil images are entirely overlapping and it can be shown that the amplitude of the lateral peaks is proportional to $A_2$. When the pupils are fully superimposed the covariance is underestimated by a factor of four. As the analysis plane is moved away from the ground the images separate on the detector leading to a more complicated, height dependent contrast adjustment. This has been addressed in detail by \cite{Avila09}. They show that using the auto--covariance of the average overlapping pupil images to normalise the individual auto--covariance can actually introduce an error of the order of tens of percent depending on the telescope aperture geometry and the altitude of the layers. This is now understood and can be corrected using theoretical approximations. However, this issue can be entirely avoided by separating the pupil images in Stereo--SCIDAR.\comment{ In this case the defocussed pupil images are fully separated and there is no loss of contrast.}

It is worth noting that the normalisation error only applies when the defocused pupil images are superimposed, which is the case for most Generalised--SCIDAR systems in use. Low-Layer SCIDAR (LOLAS, \citet{Avila08} is a variation of the SCIDAR method in which the pupils are also entirely separated, but still on one detector, and so also negates the aforementioned issue. High vertical-resolution Generalised--SCIDAR \citep{Masciadri10} also bypasses this normalisation issue by re-distributing the measured turbulence strength into discrete altitudes defined in the spatio-temporal auto--covariance.

\subsubsection{Improved target magnitude difference range}
The stellar magnitude difference for the targets for generalised--SCIDAR when the defocused images are superimposed is limited to $\sim$2.5 magnitudes \citep{Garcia_lorenzo11} at most and often only one magnitude \citep{Masciadri10}. 

The equivalent $A_2$ value for Stereo--SCIDAR is 1.0 (section~\ref{sect:subtheory}, equation~\ref{eq:sscidar_respfn}) and is independent of magnitude difference, $\Delta m$, of the target stars. Stereo--SCIDAR is limited only by signal. This means that the signal-to-noise ratio is independent of the magnitude difference. Therefore, larger stellar magnitude differences can be tolerated and hence a greater number of targets are available. This is particularly important on smaller telescopes. \changed{For example on the JKT (1~m, La Palma), using single camera SCIDAR we have a gap in the right accession (RA) angle of 5 hours in which there are no suitable targets ($\Delta m < 1.5$). Using Stereo-SCIDAR we have valid targets for all RA angles.}

\begin{figure}
	\centering
	 \includegraphics[width=0.45\textwidth]{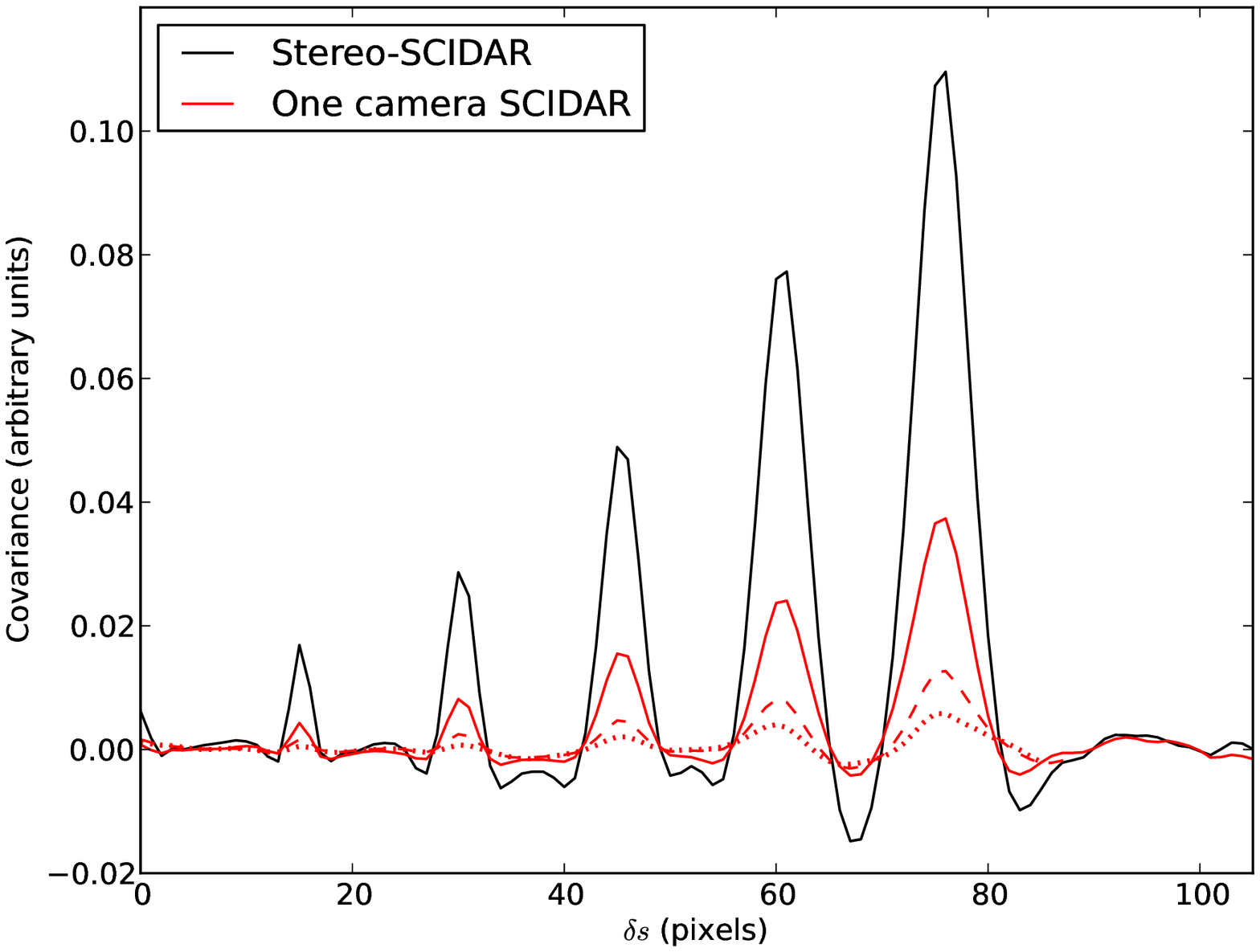} \\
	 \caption{1D cut of simulated covariance functions for Stereo--SCIDAR and single camera SCIDAR. The solid lines are for the case where the two target stars have the same magnitude, dashed lines indicate a 2 magnitude difference ($m_1$=4, $m_2$=6) and the dotted lines indicate a 3 magnitude difference in brightness ($m_1$=4, $m_2$=7).  Note that for the Stereo--SCIDAR case the three lines are co-incident. There is no loss of contrast for increasing magnitude difference.}
	 \label{fig:dm_comp}
\end{figure}

Figure~\ref{fig:dm_comp} shows the 1D cut of simulated cross--covariance functions for Stereo--SCIDAR and for conventional Generalised--SCIDAR. The simulation was for a 2.5~m telescope, 30 arcsec target separation, -3000~m conjugate altitude and one minute of data. We show the covariance function for equal magnitude stars and for targets with 2 and 3 magnitudes difference in brightness. The generalised--SCIDAR covariance peaks becomes smaller and hence increasingly more noisy for larger magnitude differences, due to contrast reduction explained in section~\ref{sect:A1A2}. The reduction in amplitude can be tolerated and included in the theoretical correction described above, however the lower signal to noise ratio is a fundamental problem and limits the possible magnitude difference of the targets for single camera Generalised--SCIDAR.

\subsubsection{Sensitivity}
\changed{
The sensitivity of the SCIDAR method is well documented in the literature \citep{Tokovinin97,Prieur01,Prieur04}. The approach taken is to calculate the sensitivity at a position $j$ in the covariance function, corresponding to a layer at altitude $h_j$, in the presence of an assumed turbulent atmosphere with a dominant layer at altitude $H_0$. 

The statistical rms noise of $B(\vec{r})$ for a single frame can be approximated by dividing the sum of the independent noise variances (scintillation, readout  and shot noise in this case, but others can be included), by the square root of the number of speckles in the overlapping area of projected pupils at altitude $h_j$. The number of speckles in the area of overlap is equal to the area of overlap of the projected pupils divided by the area of the dominant speckle size, 
\begin{equation}
A_F = \frac{f(r/D)\pi (D/2)^2}{ \pi r_F^2}, 
\end{equation}
where $f(x) = \arccos{(x)} - x(1-x^2)^{1/2}$ is the fraction of overlapping area of two full disks separated by a distance $x$ in diameter units where, $x=r/D$ and $r_F = \sqrt{\lambda H_0}/2$ is the Fresnel zone radius of the most significant speckle scale, i.e. the speckle with the largest $\sigma^2_I$, from a layer at altitude $H_0$ \citep{Tokovinin97}. Therefore, 
\begin{equation}
\Delta B(r) = \frac{(A_1B(0) + (R/N_\gamma)^2 + 1/N_\gamma) }{ \sqrt{\frac{D^2f(r)}{\lambda H_0}}},
\label{eq:rms_noise}
\end{equation}
where $A_1B(0)$ is the scintillation variance, $\sigma^2_I$, and is equal to the amplitude of the central peak in the auto-covariance function, $R$ is the rms readout noise per scintillation speckle and $N_\gamma$ is the number of photons received during the exposure per scintillation speckle. \comment{This is an approximation as it is likely that there will be several turbulent layers in the atmosphere and hence several contributions to the speckles in the analysis plane. }

This can be converted into the sensitivity in the turbulence profile by dividing by the scintillation variance of a layer with unit strength at altitude $h_j$ ($\sigma_{I,j}^2 = 19.12 \lambda^{-7/6} h_{j}^{5/6}$), using,
\begin{gather}
\Delta \tilde{J}_j =\\
\frac{5.23\times10^{-2} \lambda^{5/3}h_j^{-5/6}H_0^{1/2}(A_1B(0) + (R/N_\gamma)^2 + 1/N_\gamma)} {A_2D\sqrt{f(x) T/\tau_c }},
\label{eq:sensitivity}
\end{gather}
where $J=\int_{h_1}^{h_2} C_n^2(h)dh$, is the integrated turbulence strength over some altitude range and has units $m^{1/3}$. To calculate the rms noise of the profile, $\Delta \tilde{J}_j$, at position $j$, for the recovered profile we must include the number of independent realisations used to generate the profile. This can be approximated by $T/\tau_c$, where $T$ is the integration time and $\tau_c$ the frame rate. 

The above equations are valid for completely overlapping pupils, i.e. $h_\mathrm{conj} = 0$, for targets of equal magnitude $A_1  = 0.5$ and $A_2=0.25$. In the extreme of fully separated pupils, as with stereo--SCIDAR and LOLAS, $A_1 = A_2 = 1$. As the sensitivity of SCIDAR is proportional to $A_1/A_2$, stereo--SCIDAR can achieve double the sensitivity. This is due to the increase in contrast in the analysis plane image and hence the covariance function.

For partially separated pupils, the amplitude of the covariance peak for a particular layer depends on the separation of the pupils in the analysis plane, $h_\mathrm{conj}\theta$, and the position of the peak in the auto-covariance, $h\theta$. \cite{Avila09} show exact expressions and approximations for the relative reduction of the covariance peaks, $\epsilon(h\theta)$. 

\comment{

We know that the auto covariance function has three contributions. For partially separated pupils, as with conventional generalised-SCIDAR, these components also separate in the covariance function. If the covariance peak of the layer is in a region where the three contributions overlap, $h\theta < h_\mathrm{max}\theta - 2h_\mathrm{conj}\theta$, then the relative reduction of the covariance peak, $\epsilon(h\theta)$, is,
\begin{equation}
\epsilon(h\theta) =  \frac{\theta h_\mathrm{conj}}{D-\theta h- \theta h_\mathrm{conj}},
\label{eqn:overlapA}
\end{equation}
In the case that the layers are completely overlapping, $h_\mathrm{conj}\theta =0$ and $\epsilon(h\theta) = 0$.

If the covariance peak of a layer is in the region where it is only overlapping with the central contribution, $h_\mathrm{max}\theta - 2h_\mathrm{conj}\theta\le h\theta < h_\mathrm{max}\theta - h_\mathrm{conj}\theta$,
\begin{equation}
\epsilon(h\theta) =\frac{A_2(D-\theta h)+(1+2A_2)\theta h_\mathrm{conj}}{(1-A_2)(D-\theta h)-(1-2A_2)\theta h_\mathrm{conj}},
\end{equation}

If the covariance peak of the layer is in a region where there is no overlap, $h_\mathrm{max}\theta - h_\mathrm{conj}\theta \le h\theta < h_\mathrm{max}$,
\begin{equation}
\epsilon(h\theta) =\frac{1-A_2}{A_2}.
\end{equation}
}

The relative amplitude of the lateral covariance peak is given by $A_2(h\theta) = A_2(1+\epsilon(h\theta))$ and the central peak is $A_1(h\theta) = A_1(1+\epsilon(h\theta)/3)$. For overlapping pupils, the coefficients $A_1(h\theta)$ and $A_2(h\theta)$ collapse to $A_1$ and $A_2$ respectively and are as shown in equations~\ref{eq:A1} and \ref{eq:A2}. For separated pupils they both converge to unity.

To calculate the sensitivity of generalised-SCIDAR with partially separated pupils we can replace $A_1$ and $A_2$ in equation~\ref{eq:sensitivity} with $A_1(h\theta)$ and $A_2(h\theta)$.

In a standard generalised--SCIDAR set-up, we would expect to conjugate the analysis plane to approximately 2~km below the ground, this generally results in a shift of $\sim$10\% of the telescope diameter in the position of the two lateral set of peaks, $h_{\mathrm{conj.}}\theta$. The sensitivity is then altitude dependent with high layers having different $A_1(h\theta)$ and $A_2(h\theta)$ coefficients to the lower layers. The actual values can be calculated using the equations found within \cite{Avila09}. Figure~\ref{fig:sensitivity} shows the sensitivity as a function of layer altitude for the three cases of completely overlapping, completely separated and 90\% overlapping pupil images in the analysis plane. For this example, $D=1$~m, $H_0$=17~km, $C_n^2 = 100^{-15}$~$\mathrm{m}^{-2/3}$ and the target stars are assumed to be the same magnitude.

\begin{figure}
	\centering
	\includegraphics[width=0.5\textwidth]{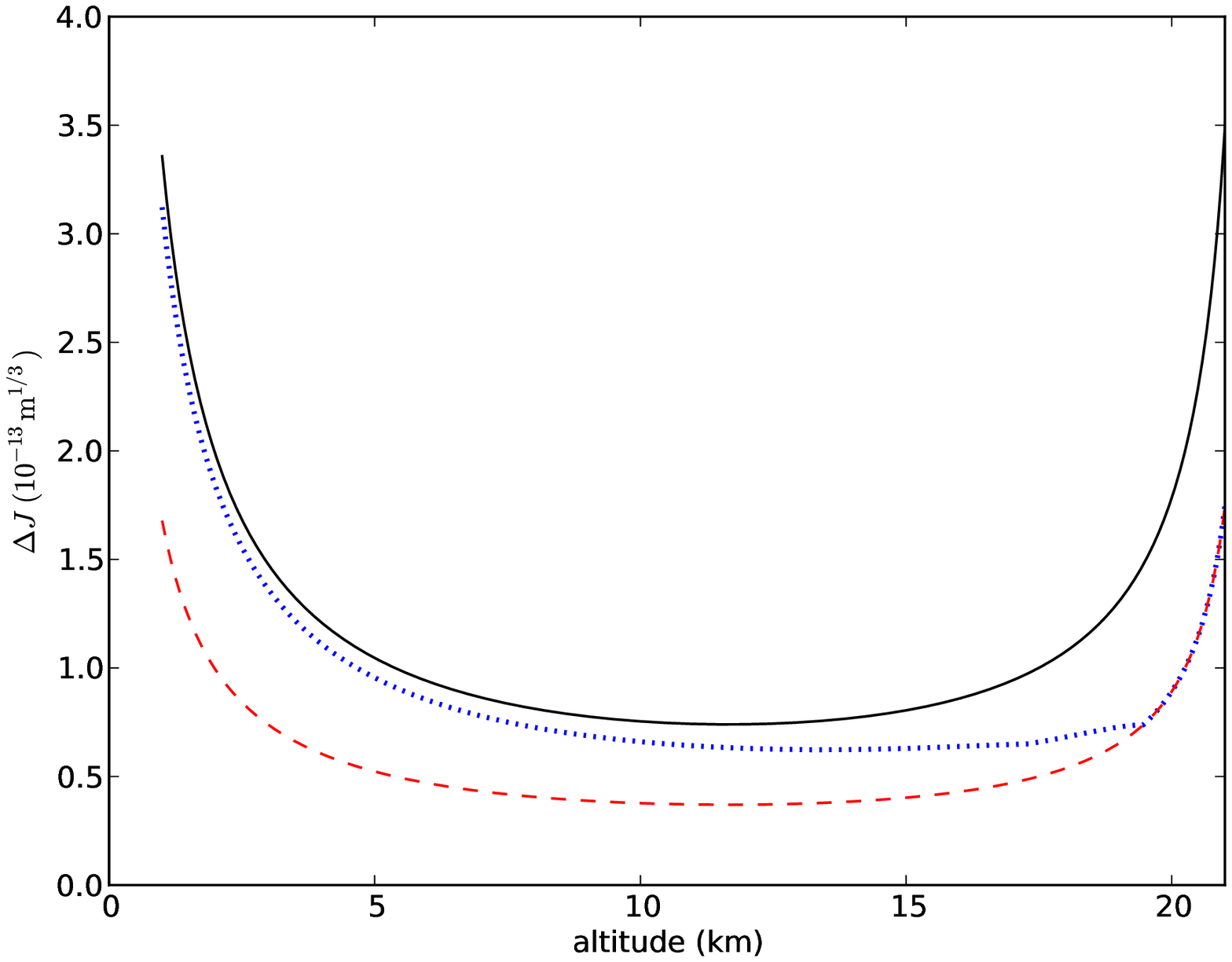}
	\caption{RMS noise of the turbulence profile, $\Delta J$, as a function of layer altitude for a single independent frame. The telescope diameter is 1~m and $\theta = 10^{\prime\prime}$. The solid line shows the completely overlapping pupil case. The dashed line is the completely separated case, with a factor of two reduction in the noise floor. The dotted line is the rms noise for the pupils with a 10\% shift (i.e. 90\% overlap). In this case low altitude layers have the sensitivity of the overlapping case and for higher layers the sensitivity converges to that of the fully separated case. The initial drop in the curves is due to the strong altitude dependance of the amplitude of the covariance peak, higher layers have larger scintillation signals and so lower noise levels. The increase in noise at higher altitudes is due to the reduction in overlap of the projected pupils, increasing the statistical noise.}
	\protect\label{fig:sensitivity}
\end{figure}

Although LOLAS also separates the pupil images of the target stars, stereo--SCIDAR still has an advantage when it comes to observing targets with a larger difference in brightness. Having two cameras permits the EM gain to be individually set for each camera, allowing for optimum gain for both targets regardless of stellar magnitude. This allows us to maximise the dynamic range for each pupil image without saturation.

}

\subsubsection{Wind velocity estimation}
\label{sect:wind_prof}
The wind velocity (speed and direction) of the strongest turbulent layers can be estimated by measuring the movement of the correlation peak of each layer in the spatio-temporal covariance function. If we assume `frozen' flow of the turbulence then the scintillation pattern will cross the telescope pupil with the same velocity as the layer. By comparing subsequent frames of the spatio-temporal covariance function the covariance peak will also move allowing  estimation of the wind velocity.

Wind velocity profiling is difficult with single camera SCIDAR. This is because each layer contributes three peaks to the covariance function which quickly become spatially confused in the spatio--temporal covariance function. This prohibits the use of geometric algorithms and instead one must identify the layers using a spatio--temporal Fourier analysis. The wavelets approach of \cite{Garcia_lorenzo06} relies on measuring the spatial frequencies inherent in the separation of the correlation peaks. \cite{Prieur04} developed an automatic algorithm based on a modified CLEAN method. Both procedures require fine parameter tuning and are computationally intensive making an automatic algorithm difficult.

For Stereo--SCIDAR we use a geometric algorithm to trace the peaks and calculate vectors between temporally adjacent frames. \changed{This is possible due to the lower noise in the covariance function and the fact that we only have one covariance peak per turbulent layer, reducing the possibility of confusion.} We calculate the spatio-temporal covariance functions with temporal delays, $\delta t$, from -3 frames up to +3 frames and the peaks are identified using a Laplacian of Gaussian \changed{filter. This involves smoothing the covariance function by convolution with a Gaussian kernel and then a Laplacian operator is used to calculate the second order spatial derivative of the 2D function, effectively selecting regions with high gradients of intensity. We then select covariance peaks by recording the co-ordinates of the brightest pixel in the covariance and then subtracting the scaled Gaussian kernel and repeating until there are no peaks above three times the standard deviation of the covariance}. This has been automated and we calculate the wind profile along with the turbulence profile for all of our data in real-time. Using the Andor Luca EMCCD cameras with windowing we attain a frame rate of approximately 90~Hz. For the instrument installed on the 1~m JKT we have 80 pixels across the pupil and we can therefore achieve a wind speed \changed{scale} of $\sim$1~ms$^{-1}$ \changed{per pixel}. However, this accuracy can be improved by centroiding the correlation peaks to sub-pixel accuracy. In order to calculate the wind velocity of a layer the covariance peak for that layer must appear in three consecutive frames of the spatio--temporal cross--covariance.  Therefore, the maximum wind velocity that can be measured with the 1~m JKT is 36~m/s and with the 2.5~m NOT is 90~m/s. The worst case is a high layer where the wind direction is such that it is travelling at an angle perpendicular to the line joining the two stars. In this case we would see the layer in the dt=0 frame but it is possible that it will not appear in any other frame. Therefore, it is possible that high layers will exist for which we can not obtain velocity estimates. \changed{This can be improved by using larger telescopes or narrower targets, increasing the maximum profiling altitude. These high layers will be seen in the spatio-temporal auto-covariance and so wind velocities can still be deduced although the altitude information would then be lost.}

It is possible to use the wind profile to further enhance the resolution of the optical turbulence profile. Two layers that are unresolved in the turbulence profile may separate in the spatio-temporal cross--covariance due to a difference in wind velocities. \cite{Egner07} show that this method can be very effective.

\section{Data Reduction}
In existing SCIDAR systems the intensity patterns from each target overlap on a single detector and the profile is retrieved from the auto--covariance function of the combined image. As each pupil image is recorded separately for Stereo--SCIDAR the images can be normalised in a way which is not possible with conventional SCIDAR. The images are independently background subtracted, offset to zero mean and scaled to have unit integrated \changed{covariance strength}. This ensures that despite differing magnitudes of the stars, one does not dominate the other in the cross--covariance. Unlike single-camera generalised--SCIDAR (using the image auto--covariance), in Stereo--SCIDAR each layer only contributes one peak in the cross--covariance. To retrieve the turbulence profile we are required to solve the inverse problem, written in matrix form,
\begin{equation}
	B_{\ast\ast}\left(\vec{r},z\right) = K_z\left(\vec{r},h\right) C_n^2\left(h\right) +n(\vec{r}),
\end{equation}
where $B_{\ast\ast}\left(\vec{r},z\right)$ is the measured scintillation cross--covariance, $K_z\left(\vec{r},h\right)$ is the matrix of Stereo--SCIDAR responses to unit turbulence at different heights, for a given conjugate altitude $z$ and $n(\vec{r})$ is the noise in the cross--covariance.
The inverse problem is solved using a non--negative least squares inversion \citep*{Lawson74}, to retrieve an estimate for the $C_n^2\left(h\right)$ profile. \changed{A cut of the cross--covariance function from the centre in the direction joining the two target stars is used as the input.}

\changed{The value of the wavelength used for simulation and data reduction is $\lambda= 500\mathrm{nm}$, which corresponds approximately to the peak of the spectral sensitivity of the detector.}

\section{Results}
\changed{Stereo-SCIDAR was operated on the JKT and NOT for a total of 25 nights between February and September 2013. The results from these observations will be analysed and presented in a future publication. Here we present a selection of interesting examples collected using this system.}

\subsection{Example profile}
\begin{figure*}
\centering
$\begin{array}{c}
\includegraphics[width=\textwidth]{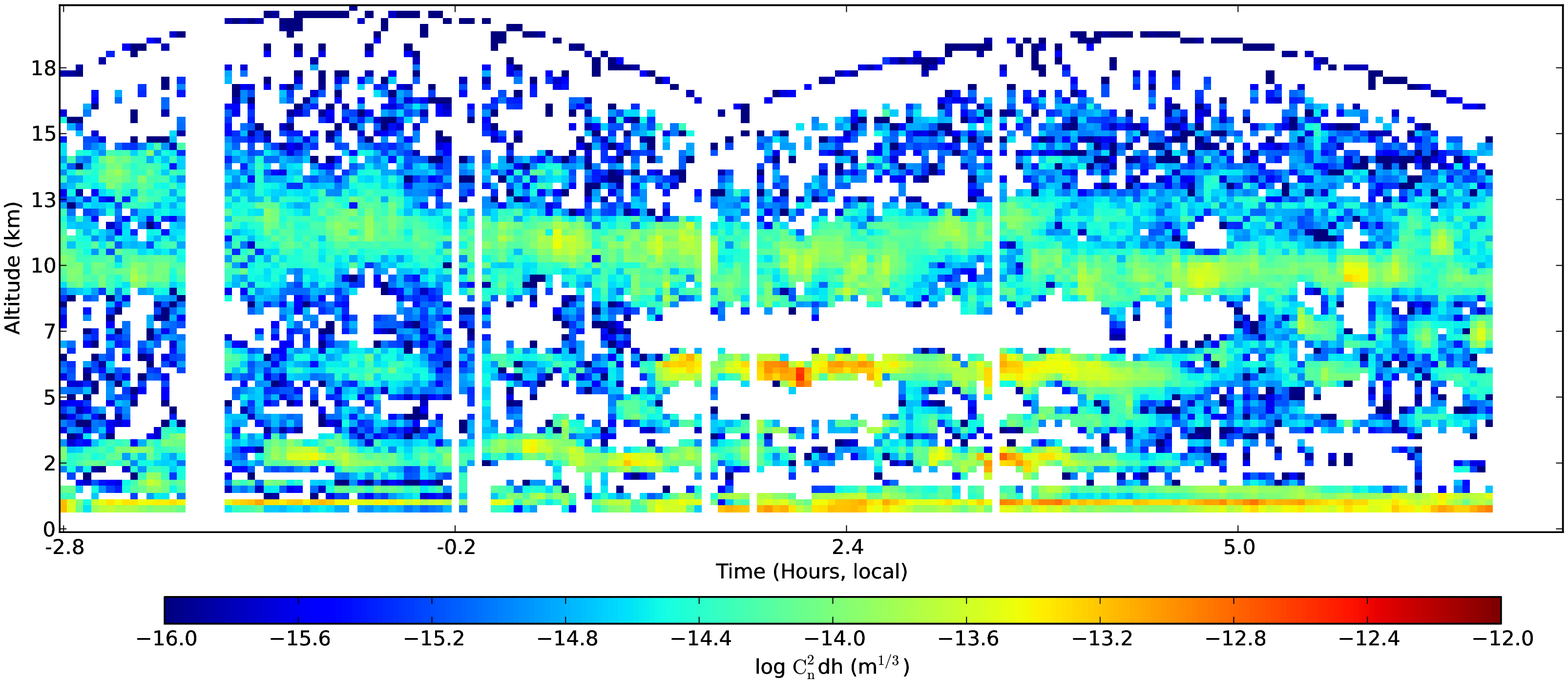}\\[0.01mm]
\includegraphics[width=\textwidth]{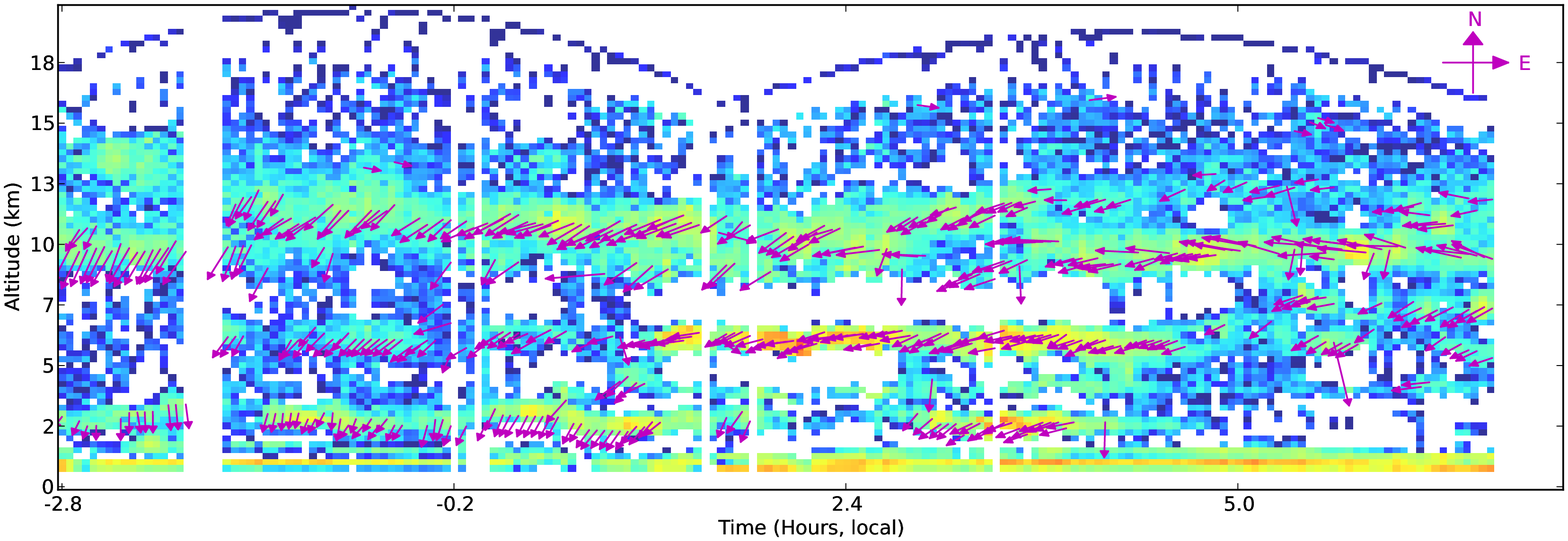}
\end{array}$
\caption{Example turbulence profile from the JKT, La Palma, 15th September 2013. The upper plot shows the profile of the optical turbulence as a function of time. The lower plot is the same but with the layer wind velocities overlaid. The length of the arrows denote the relative wind speed and the direction corresponds to the turbulent layer direction as defined by the cardinal directions shown in the top right of the lower plot. The conjugate altitude \changed{of the analysis plane }was set to 0~m.}
\label{fig:eg_prof_20130524}
\end{figure*}

Figure~\ref{fig:eg_prof_20130524} shows the turbulence profile from the JKT, recorded on the night of the 15th/16th September, 2013.\changed{ The analysis plane was set to 0~m to remove any contribution from the dome and ground and hence increase the SNR of the higher layers. Two targets were observed during the night, the first had a stellar magnitude difference of $\Delta_m = 0.7$, the second $\Delta m = 2.7$, both had an angular separation of $\sim10^{\prime\prime}$.} The upper plot shows only the optical turbulence profile. The wind profiles are overlaid on a separate plot for clarity. We do not analyse the profiles in any way here but simply point out a few interesting features that we have seen in the data collected to date. Qualitatively, we see several branching points where a turbulent layer seems to split into two layers and diverge in altitude. We also see the temporal evolution of these turbulent layers, which often seem to be correlated in altitude. The line delineating the maximum altitude is an artefact in the cross-covariance function.

\subsection{Wind velocity profiling}
\begin{figure*}
\centering
$\begin{array}{ccccc}
	\includegraphics[width=0.18\textwidth]{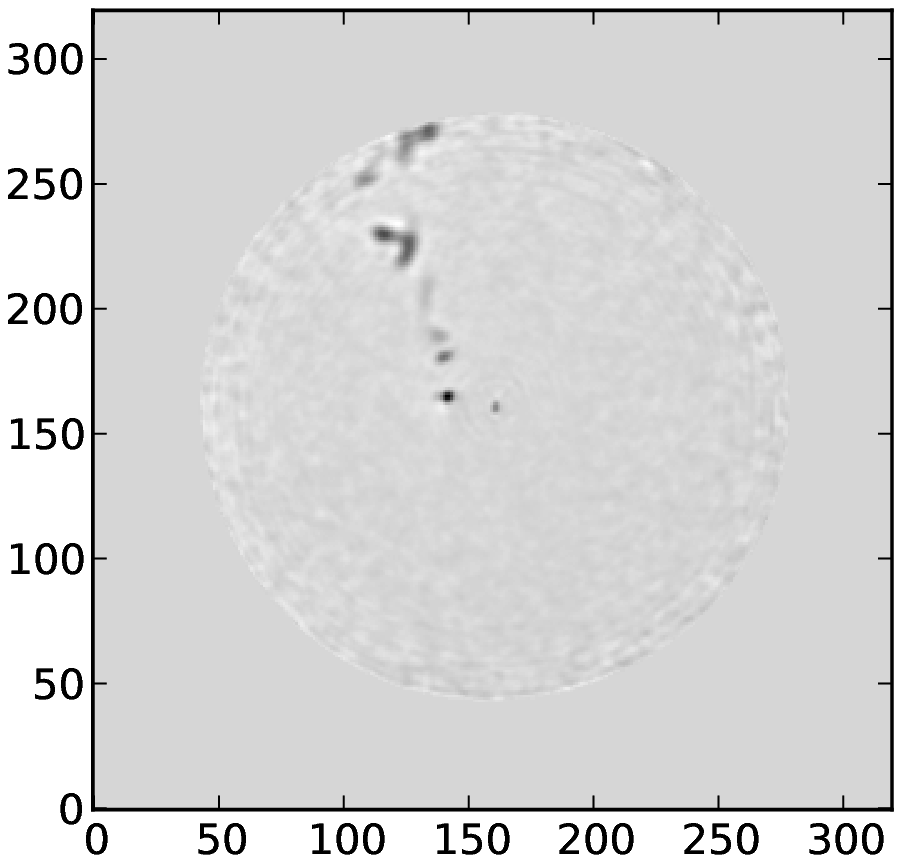}&
	\includegraphics[width=0.18\textwidth]{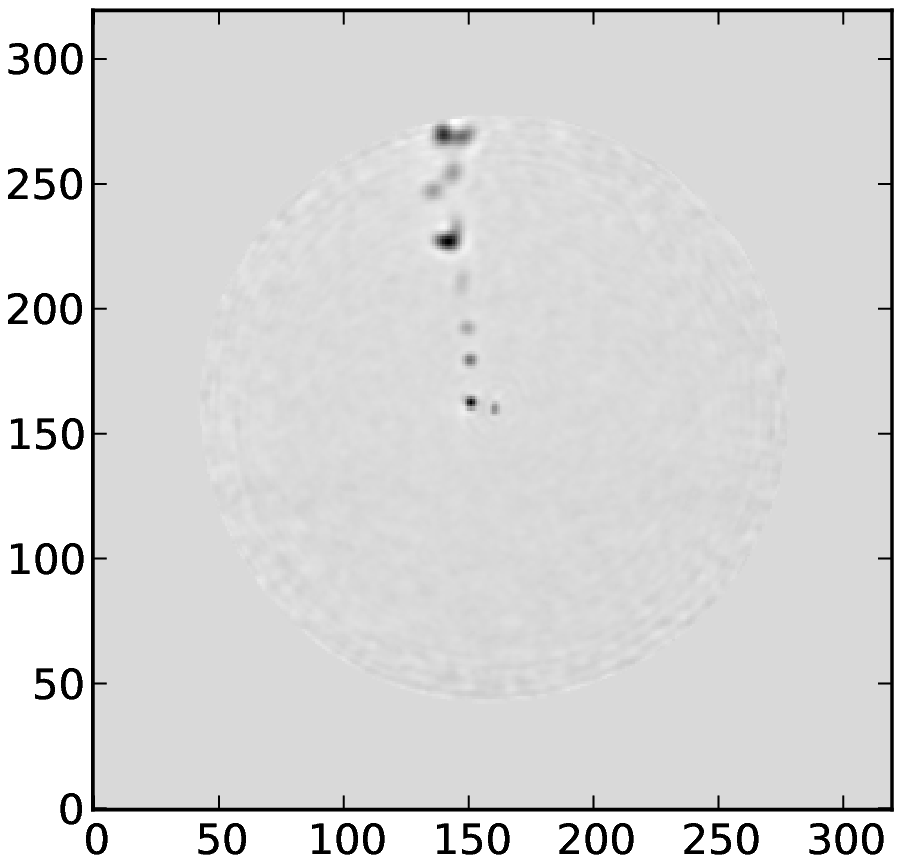}&
	\includegraphics[width=0.18\textwidth]{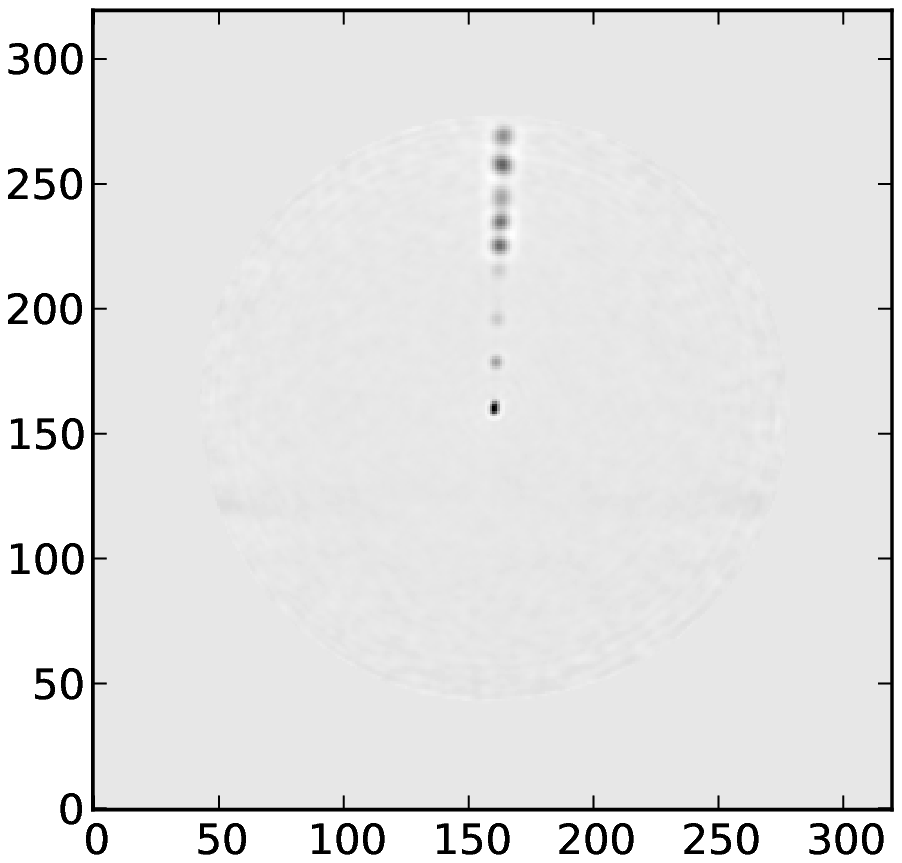}&
	\includegraphics[width=0.18\textwidth]{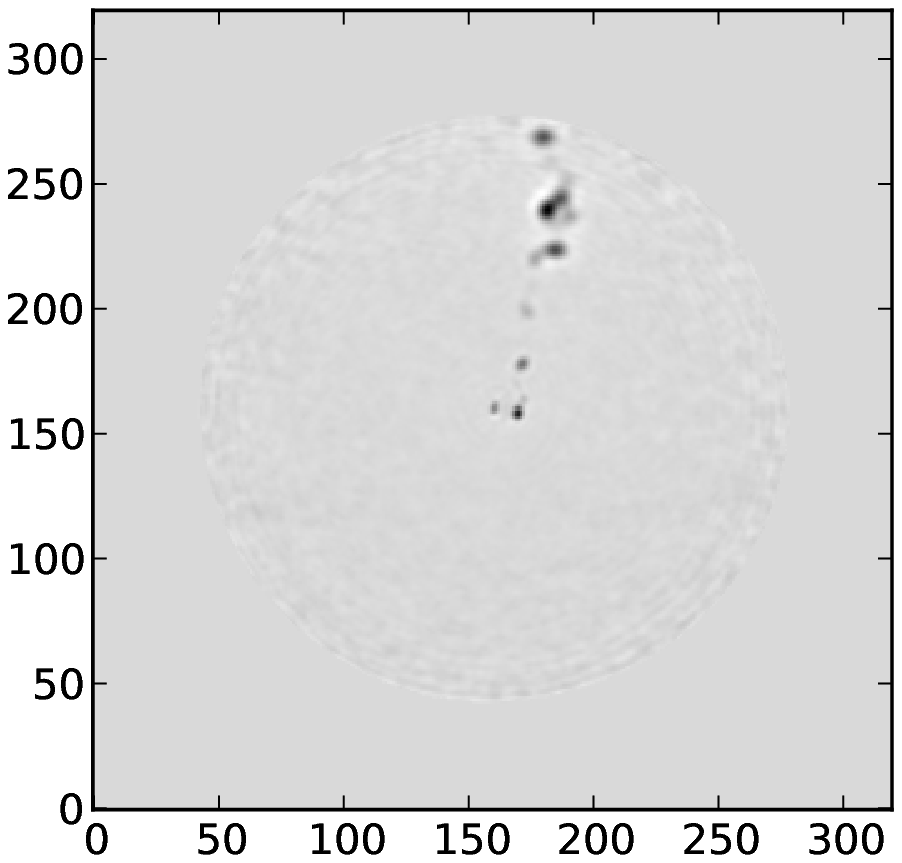}&
	\includegraphics[width=0.18\textwidth]{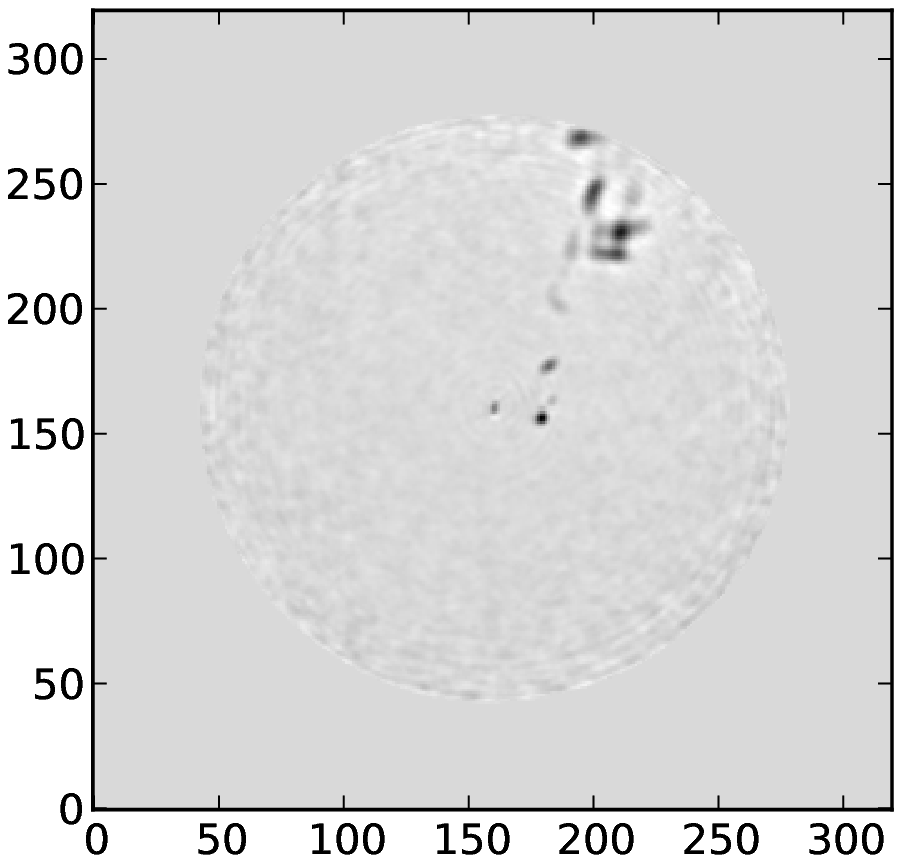}\\
	\mbox{\bf{(a)}}  & \mbox{\bf{(b)}} &\mbox{\bf{(c)}}  & \mbox{\bf{(d)}} & \mbox{\bf{(e)}}  	
\end{array}$
\caption{Spatio--temporal cross--covariance functions for the data taken at a conjugate altitude of -2~km (intensity scale inverted for clarity). The plots show cross--covariance functions generated with temporal delays equal to 1 frame ($\sim$10~ms) from -2 frames (a) to +2 frames (e). The case of no temporal delay is shown in (c). By examining the position of these peaks in subsequent frames the wind velocity (magnitude and direction) can be calculated.}
	\protect\label{fig:onskywind}
\end{figure*}

\begin{figure}
	\centering
	\includegraphics[width=0.45\textwidth]{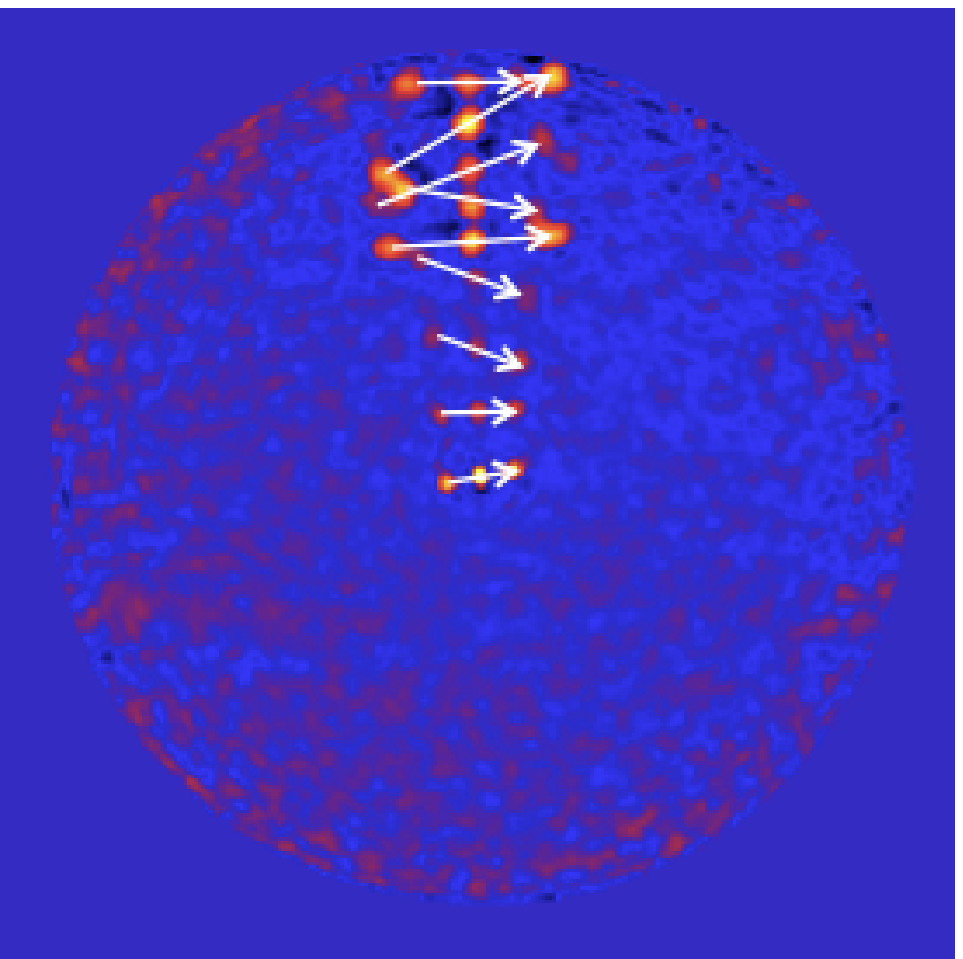}
	\caption{The sum of three consecutive spatio-temporal cross--covariance frames. We show the sum on one image to demonstrate the wind velocity estimation process. The arrows indicate the detected layers and velocities.}
	\protect\label{fig:xcov_sum3}
\end{figure}

Figure~\ref{fig:onskywind} shows the spatio-temporal cross--covariance functions for delays in the range -2 frames up to +2 frames.
If we add the central three frames together (b, c and d from figure~\ref{fig:onskywind}) then it becomes easier to see the velocity of each layer (figure~\ref{fig:xcov_sum3}). To build a wind profile we assume frozen flow and implement a geometric algorithm. We make a least squares fit between equi-spaced peaks in adjacent frames. We then do the same for several sets of three frames (positive and negative temporal offsets) so that we can detect layers even if they leave the scope of the cross--covariance function. To detect the velocity of a layer we require it to be seen in at least one set of three frames.\comment{, reduceing the probability of false positive detections.}

\changed{
Figure~\ref{fig:wind_speed_dist} shows an example distribution of wind speeds as a function of altitude for 160 profiles taken throughout the night of 13th September 2013, from the JKT, La Palma. The dashed line indicates the maximum wind speed that can be measured with the current instrument on the 1~m JKT, with a framerate of $\sim$72~Hz, this is 36~m/s (section~\ref{sect:wind_prof}).
\begin{figure}
	\centering
	\includegraphics[width=0.5\textwidth]{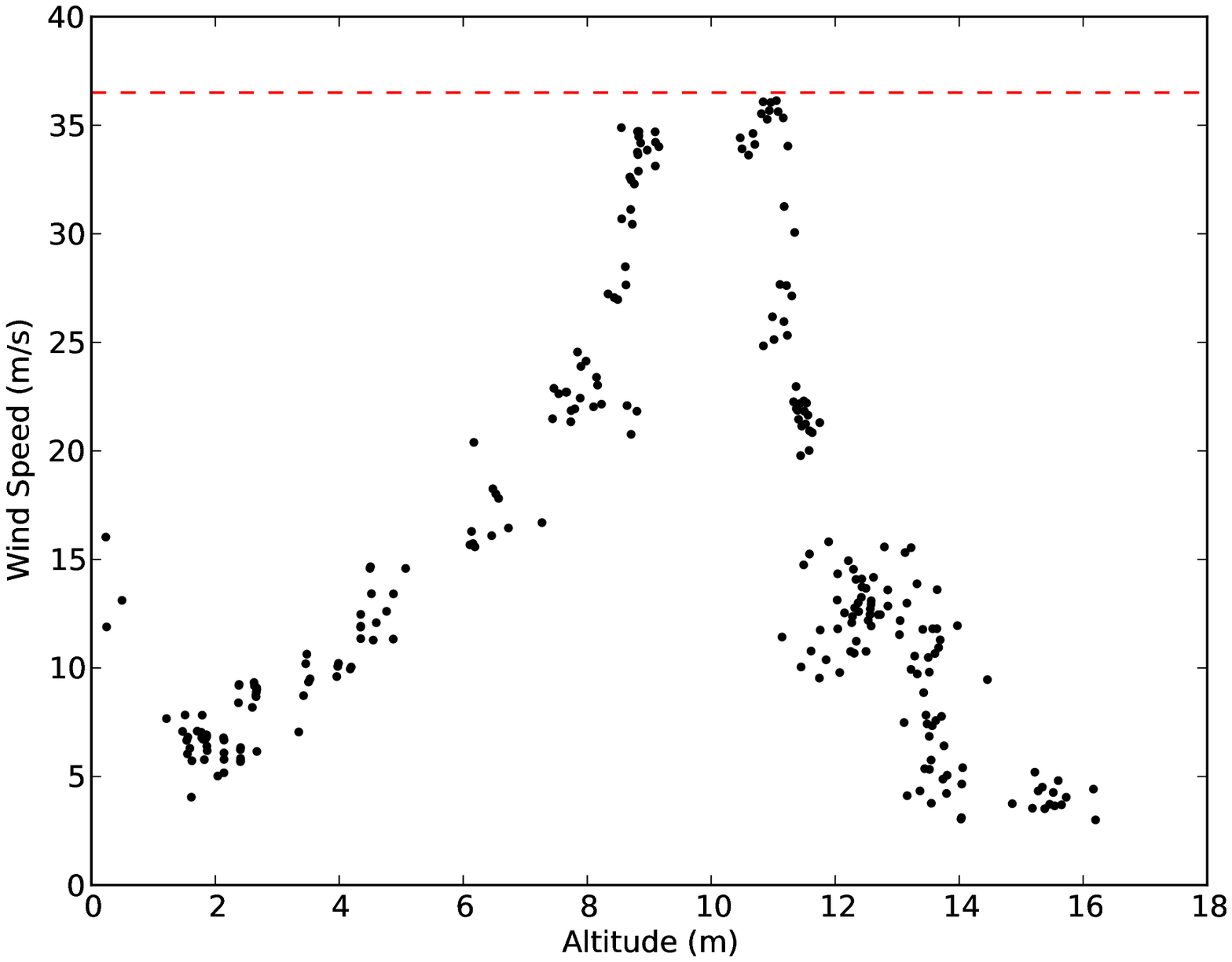}
	\caption{Example distribution of wind speeds as a function of altitude for the night of 13th September 2013, JKT, La Palma. The dashed line indicates the maximum wind speed that can be measured with the current instrument on the JKT. The tropopause can clearly be identified at an altitude of $\sim$10~km.}
	\protect\label{fig:wind_speed_dist}
\end{figure}
}

\subsection{Dome seeing estimates}
We are able to estimate and remove the contribution of the dome seeing from the covariance function using an extension of the method explained in \cite{Avila98}. For dome seeing estimates the conjugate altitude of the analysis plane must be a few kilometres away from the telescope pupil. Any turbulent layer in the atmosphere will be blown across the pupil of the telescope with some wind speed. Dome seeing will develop slowly and will therefore remain as a peak in the spatio-temporal covariance even with several frames temporal offset. The amplitude of this central peak will slowly decay as the dome seeing decorrelates.

By looking at the amplitude of this central peak as a function of time we will see that only the first few frames will also include the contribution from the surface turbulent layer which will move away from the central position with the velocity of the wind. We can then extrapolate back to zero offset covariance and estimate a value for the covariance value of the dome seeing. Using this technique we are able to remove the dome seeing contribution from the surface turbulent layer strength. 

\begin{figure}
	\centering
	\includegraphics[width=0.45\textwidth]{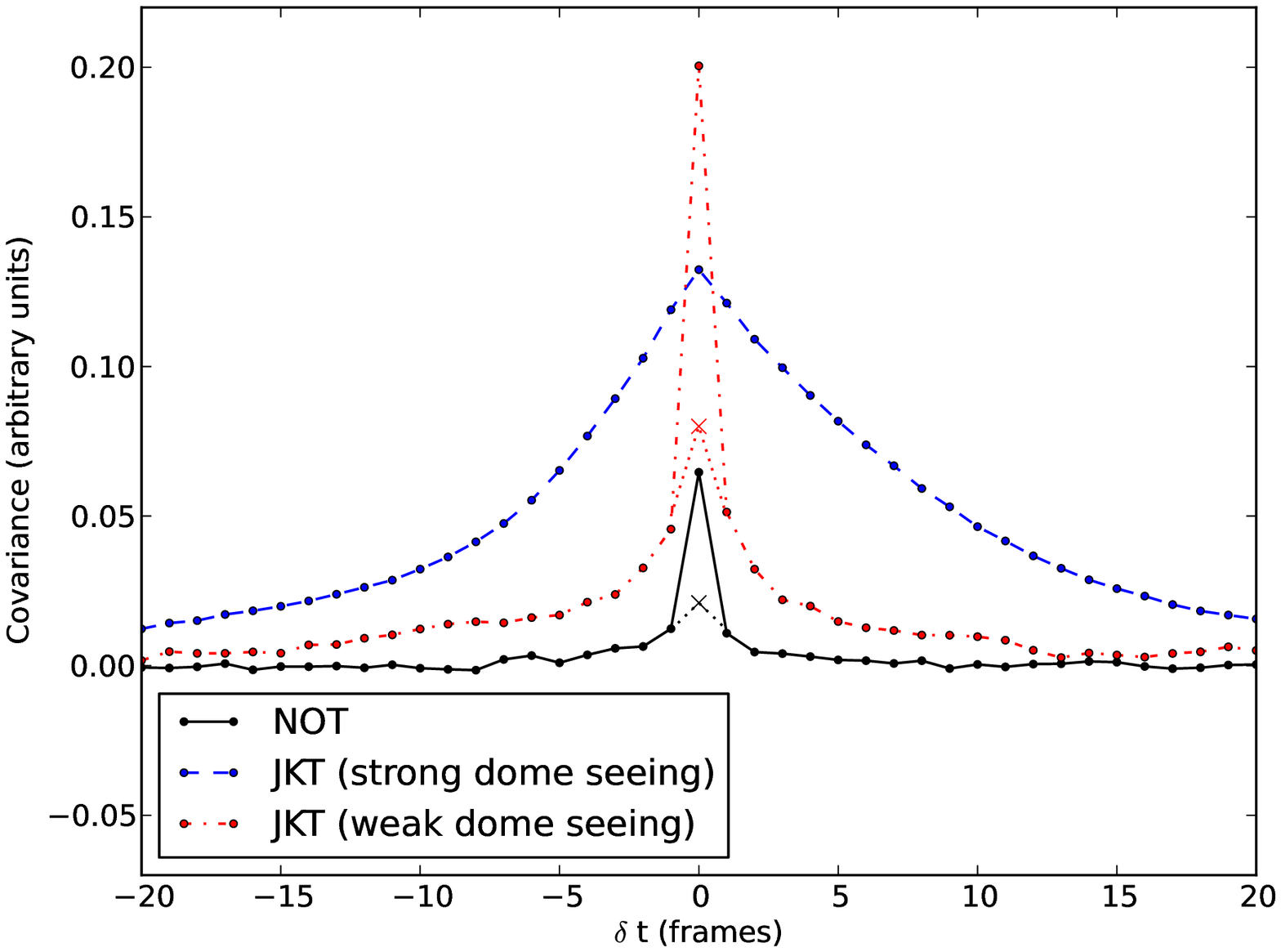}
	\caption{The central value of the auto--covariance as a function of temporal delay between the two images. The central spike visible in two of the three curves corresponds to the strength of the external surface layer turbulence. This moves away from the centre of the auto--covariance with a velocity given by the wind velocity. The remnant covariance that lingers in the central peak is due to the seeing in the dome of the telescope and decorrelates slowly. The solid curve is for the NOT, anecdotally known for its low dome seeing. We see here that the remnant is extremely small. The dash-dot JKT curve was observed at a time when we did not experience much dome seeing, we do however observe a strong surface layer. By extrapolating the decay of the remnant covariance back to the centre we can estimate the covariance due to the dome seeing (marked by a dotted line and a cross). The dashed curve demonstrates a scenario when we observed very little external turbulence and very strong dome seeing. This was at a time when that outside windspeed was effectively zero and shows that the SCIDAR, as with many remote sensing instruments, fails in these conditions.}
	\protect\label{fig:dome_seeing}
\end{figure}

Here we use the spatio-temporal auto--covariance function (the covariance of each pupil image with itself). In the auto-covariance function any altitude information is removed and all layers appear as covariance peaks at the centre of the covariance function. They then move radially away from the central point in the spatio-temporal auto-covariance, with a velocity determined by the velocity of the turbulent layer. The auto-covariance is more useful than the cross-covariance for this analysis, as in the spatio-temporal cross--covariance high altitude layers can move through the central peak, adding noise to the dome seeing estimate. As we are only using the auto-covariance this method could be utilised while observing a single star and thus increase the number of targets available. This would also allow us to use a smaller telescope dedicated for dome seeing measurements. \comment{However, it is difficult to convert the dome covariance into a seeing value as the dome turbulent is not necessarily Kolmogorov and will probably have a small $L_0$. }Figure~\ref{fig:dome_seeing} shows measurements of the dome seeing at both the JKT and NOT telescopes. 

\section{Conclusion}
We have developed and tested a new generalised-SCIDAR remote sensing instrument for the characterisation of optical turbulence above astronomical sites. In this technique, which we refer to as stereo--SCIDAR, the light from each component of a target double star is imaged on a separate detector. Separating the light from each star allows us to avoid the normalisation issue of generalised--SCIDAR, and increases the useable magnitude difference of the targets, resulting in 100\% time coverage for La Palma on a 1~m telescope. It also achieves an increase in the signal-to-noise ratio of a factor of 2 for a target pair of stars of the same magnitude, a factor of 3 for $\Delta m = 1$ and a factor of 6.3 if there is two magnitudes difference in the brightness of the targets. We have successfully used stereo--SCIDAR with 2.7 magnitudes difference in the target brightness. This yields an increase of a factor of 12 in the sensitivity over single camera SCIDAR. Stereo--SCIDAR also provides a simple, automatic, technique for the detection of the velocity of the atmospheric turbulent layers and the dome seeing.

A limited on--sky test demonstrated the key concepts of the technique. We show several examples from the on-sky data, including an example turbulence profile complete with wind velocity measurements and estimates of the dome seeing for the NOT and JKT.

\section*{Acknowledgments}

We are grateful to the Science and Technology Facilities Committee (STFC) for financial support (grant reference ST/J001236/1). FP7/2013-2016: The research leading to these results has received funding from the European Community's Seventh Framework Programme (FP7/2013-2016) under grant agreement number 312430 (OPTICON). The Jacobus Kapteyn Telescope is operated on the island of La Palma by the Isaac Newton Group in the Spanish Observatorio del Roque de los Muchachos of the Instituto de Astrof'sica de Canarias. Some data was based on observations made with the Nordic Optical Telescope, operated by the Nordic Optical Telescope Scientific Association at the Observatorio del Roque de los Muchachos, La Palma, Spain, of the Instituto de Astrofisica de Canarias. R.A. acknowledges funding provided by grant IN115013 from PAPPIT DGAPA-UNAM. The authors would like to thank E.~Masciadri for her very helpful comments during the preparation of this paper. Data used in the preparation of this paper is available on request from the authors. We thank the reviewer, A.~Tokovinin, for his insightful comments which certainly led to a clearer paper.

\bibliographystyle{mn2e}

\end{document}